VIEWPOINT

# Governed Collaborative Memory as Artificial Selection in LLM-Based Multi-Agent Systems

Diego F. Cuadros[1, 2, 3, *] | Abdoul-Aziz Maiga[4] | Helen Meskhidze[3, 5] | Andre Curtis-Trudel[3, 5]

[1] Digital Epidemiology Laboratory, Digital Futures, University of Cincinnati, Cincinnati, OH, USA
[2] Department of Biological Sciences, University of Cincinnati, Cincinnati, OH, USA
[3] Center for Humanities and Technology (CHaT), University of Cincinnati, Cincinnati, OH, USA
[4] Norwegian University of Science and Technology, Trondheim, Norway
[5] Department of Philosophy, University of Cincinnati, Cincinnati, OH, USA
[*] Correspondence: diego.cuadros@uc.edu

## ABSTRACT

Persistent memory is turning language-model-based agents from stateless participants in isolated interactions into state-bearing components of LLM-based multi-agent systems. As memory becomes durable, reloadable, and behavior-shaping across agents, sessions, or versions, a design question arises that is not captured by retrieval accuracy or access control alone: which candidate memories should become shared institutional state? This Viewpoint frames that problem as governed collaborative memory. We argue that memory governance functions as a selection regime, determining which memory variants persist, which remain private, and which are rejected, abstained from, or superseded. We distinguish ungoverned persistence, constitutional or hybrid selection, automatic metric-based selection, and human-ratified artificial selection, emphasizing that these regimes are not a ranking but a design choice over target properties. We then describe a layered architecture that separates agent-local memory, shared institutional memory, archive memory, and project-continuity memory, with provenance and version lineage making selection inspectable. Documented traces from one running LLM-based multi-agent ecosystem illustrate unmanaged false-memory persistence, ratified institutional memory, rejection and revision, identity-preserving expansion, and governance-as-learning. The contribution is a design agenda: persistent LLM-based multi-agent systems should evaluate memory not only for recall and performance, but also for provenance fidelity, selection traceability, epistemic quality, correction pathways, and role preservation.

### DESIGN CLAIMS

- Persistent LLM-based multi-agent systems need explicit selection regimes for shared memory.
- Governed collaborative memory separates agent-local memory, shared institutional memory, archive memory, and project-continuity memory.

- Human-ratified artificial selection is useful when target properties include judgment quality, provenance fidelity, institutional coherence, and role preservation.
- Rejected, abstained, superseded, and ratified memories should all remain inspectable selection outcomes.
- The documented traces illustrate the design pattern but do not establish causal superiority over automatic or ungoverned regimes.

## 1 Agent Memory Is Becoming Shared State

Multi-agent systems increasingly include language-model-based agents that act across repeated interactions, tool calls, documents, and collaborations. In such systems, memory is no longer only a local convenience for dialogue continuity. It can become part of the system state that determines how agents coordinate, what they treat as known, which failures recur, and which lessons persist. Persistent memory turns an agent from a stateless participant in a single interaction into a state-bearing component of a larger system.

This shift matters because some agent memories are becoming heritable. We use the term operationally, not biologically: memory becomes heritable when it is durable enough to be reloaded, influential enough to shape behavior, and portable enough to affect future agents, sessions, or system versions. A note saved after one interaction can later guide another agent's decision; a lesson extracted from one failure can become a rule for future work; a memory update can change how a system behaves without changing model weights. Once this occurs, the central question is no longer only how agents remember, but which memories should be allowed to shape future behavior.

Multi-agent systems have long traditions of work on agent state, coordination, and shared resources [1]. Our focus is narrower: LLM-based multi-agent systems in which natural-language memories can be written, summarized, shared, reloaded, and revised across sessions. Existing work has made substantial progress on this newer memory problem. Agent-memory research now includes broad taxonomies of memory mechanisms, memory-stream architectures for believable agents, and production systems for scalable long-term memory [2], [3], [4]. This work establishes that memory is becoming a first-class resource for autonomous agents. It also shows that memory is difficult: agents must decide what to store, how to retrieve it, how to update it, when to forget it, and how to evaluate whether memory improves downstream behavior.

The multi-agent setting adds another layer. Recent systems and frameworks address private and shared memory tiers, dynamic access control, provenance, memory consistency, semantic drift, privacy, and role-aligned agent memory [5], [6], [7]. These are essential advances. Access control asks who can read or write a memory. Provenance asks where a memory came from. Safety-oriented governance asks how to prevent contamination, leakage, or inconsistency. Role-aligned memory asks how agents can preserve specialized perspectives. Combining access control with safety governance still leaves a further design question underdeveloped: which candidate memories should become shared, durable, and behavior-shaping across a heterogeneous agent ecosystem without eroding role-specific identity?

This Viewpoint argues that the missing question is not who can read a memory, or how accurately it can be retrieved, but who or what decides that a candidate memory should become shared institutional state. We call this problem governed collaborative memory: deciding which candidate memories become shared, durable, and behavior-shaping institutional state. The argument is grounded in recent work on agent memory, shared memory, and memory governance, and in documented traces from one running LLM-based multi-agent system. The claim is deliberately narrow: memory governance is not merely access control, summarization, or retrieval policy. In persistent LLM-based multi-agent systems, governance also functions as a selection regime. Agents generate candidate memory variants, governance filters them, and selected memories become durable shared state while other memories remain private, transient, rejected, or superseded.

# 2 Selection Regimes for Shared Memory

A selection regime is the policy, mechanism, or governance process that determines which candidate memory variants persist long enough to shape future system behavior. The term is useful here as a design abstraction, not as a claim that agents are biological organisms. Any persistent memory system must answer a selection question, even if it answers implicitly. If an agent writes every plausible summary into durable memory, that is one regime. If memories persist only after automated tests, human-authored principles, or explicit review, those are different regimes. Each creates different system-level dynamics.

The least governed regime is ungoverned persistence: candidate memories enter durable storage without deliberate evaluation. This is simple and low-friction, but it creates a direct path from plausible error to persistent state. In memory-bearing agents, this failure mode is not just hallucination at one moment; it is falsehood that can be reloaded, repeated, and propagated. Automatic selection addresses a different part of the problem. In self-evolving agent systems, proposed changes can be assessed by benchmarks, tests, reward signals, or other measurable criteria before being committed to future behavior [8]. This is powerful when the relevant target is measurable. It is less reliable when the target is judgment quality, epistemic honesty, role diversity, or institutional coherence.

Between ungoverned persistence and direct human ratification lies a constitutional or hybrid regime. Here, humans specify principles, constraints, or reward models, but enforcement is delegated to automated processes. RLHF and Constitutional AI illustrate this pattern: human preference or principle design shapes the selection environment, while automated evaluation or self-critique applies it at scale [9]. A fourth regime is human-ratified artificial selection, in which an operator or governance process directly determines whether a candidate memory becomes durable shared state. This regime is slower and less scalable, but it can select for qualities that do not have clean metrics.

The point is not that one regime is universally superior. The design question is not whether selection should be automated, but which traits can be safely selected automatically and which require explicit judgment.

| Regime | Mechanism | Example | Selects for | Cannot select for | Failure mode |
|---|---|---|---|---|---|
| Ungoverned persistence | Candidate memories persist without deliberate evaluation | Ungoverned auto-memory | Low-friction persistence | Quality, truth, provenance | Drift, false-memory persistence |
| Constitutional / hybrid | Human-authored principles enforced automatically | RLHF, Constitutional AI | Scalable principle compliance | Traits omitted from principles | Principle gaps, reward hacking |
| Automatic | Metrics or tests decide persistence | Autogenesis-style self-evolution | Measurable performance | Judgment, coherence, epistemic honesty, diversity | Goodhart effects, local optima |
| Human-ratified artificial | Operator or governance process decides persistence | Governed evidence-and-ratification protocol | Judgment quality, institutional coherence, role diversity | Scale-dependent traits the operator cannot evaluate | Bottleneck, operator bias |

**Table 1 |** Selection regimes for shared memory in LLM-based multi-agent systems. The regimes are design choices over target properties, not a ranking from worst to best.

Shared memory also creates an identity-sharing tradeoff. In heterogeneous multi-agent systems, agent identity should be understood functionally: the stable role, constraints, priors, tools, behavioral tendencies, and failure modes that make one agent useful in a different way from another. Private identity memory matters because role specialization is part of the value of a multi-agent system. For example, an architecture reviewer, an operations synthesizer, an archive-retrieval agent, and a formal critic should not converge into the same generic assistant simply because they share the same institutional memory.

At the same time, too little sharing prevents institutional learning. If every agent learns only locally, the system repeats errors, loses cross-agent discoveries, and cannot build durable coordination norms. Too much sharing creates the opposite failure mode: homogenization, correlated mistakes, over-propagation of false memories, and loss of cognitive diversity. The problem is not simply to maximize sharing; it is to decide which memories should cross the boundary between private role-specific identity and shared institutional state.

This framing is consistent with evolutionary accounts that distinguish human-imposed fitness regimes from more open-ended ecosystem dynamics, and with philosophical accounts that distin-guish domesticated from more autonomous or “feral” AI evolution [10], [11]. But the substrate here is engineered memory governance in multi-agent systems. The question is practical: how should designers specify what varies, who or what selects, what becomes fixed, and what remains private?

The remaining design question is how to implement such selection without collapsing the heterogeneous identities that make multi-agent systems useful.

# 3 Architecture and Mechanism

If selection regimes specify what becomes durable shared state, the architectural problem is to decide where candidate memory lives, how it crosses boundaries, and how accepted memory remains inspectable. A governed-memory architecture must therefore separate two functions that are often conflated: preserving the local memory that makes heterogeneous agents useful, and maintaining the shared memory that lets the system learn as a system. Governed collaborative memory addresses this problem by treating memory stores not as one undifferentiated repository, but as separable layers with different selection burdens.

The first layer is agent-local memory. Its role is to preserve the stable constraints, heuristics, tools, and failure modes that make one agent useful in a different way from another. If all useful discoveries are pushed immediately into a common memory, role specialization can erode and agents can converge toward the same generic behavior. Agent-local memory is therefore private by default, with promotion to shared memory only when the candidate record has cross-agent relevance. The second layer is shared institutional memory: a durable store for events, principles, decisions, protocol changes, and lessons that should affect more than one agent. Without this layer, agents repeat each other’s mistakes and discoveries remain siloed. Because shared institutional memory is behavior-shaping across the ecosystem, it carries the strongest governance burden.

A third layer is archive memory: long-horizon research artifacts, historical records, and background knowledge that should remain retrievable without being treated as active institutional memory. This distinction prevents governance overload. A system should not have to ratify every historical artifact before retrieval, but it should preserve provenance and distinguish archived context from ratified shared lessons. A fourth layer is project continuity memory: active task state, handoff context, and short-term coordination information. This prevents scope contamination between temporary project state and durable institutional knowledge. The architecture is not a hierarchy of importance; it is a separation of governance burdens.

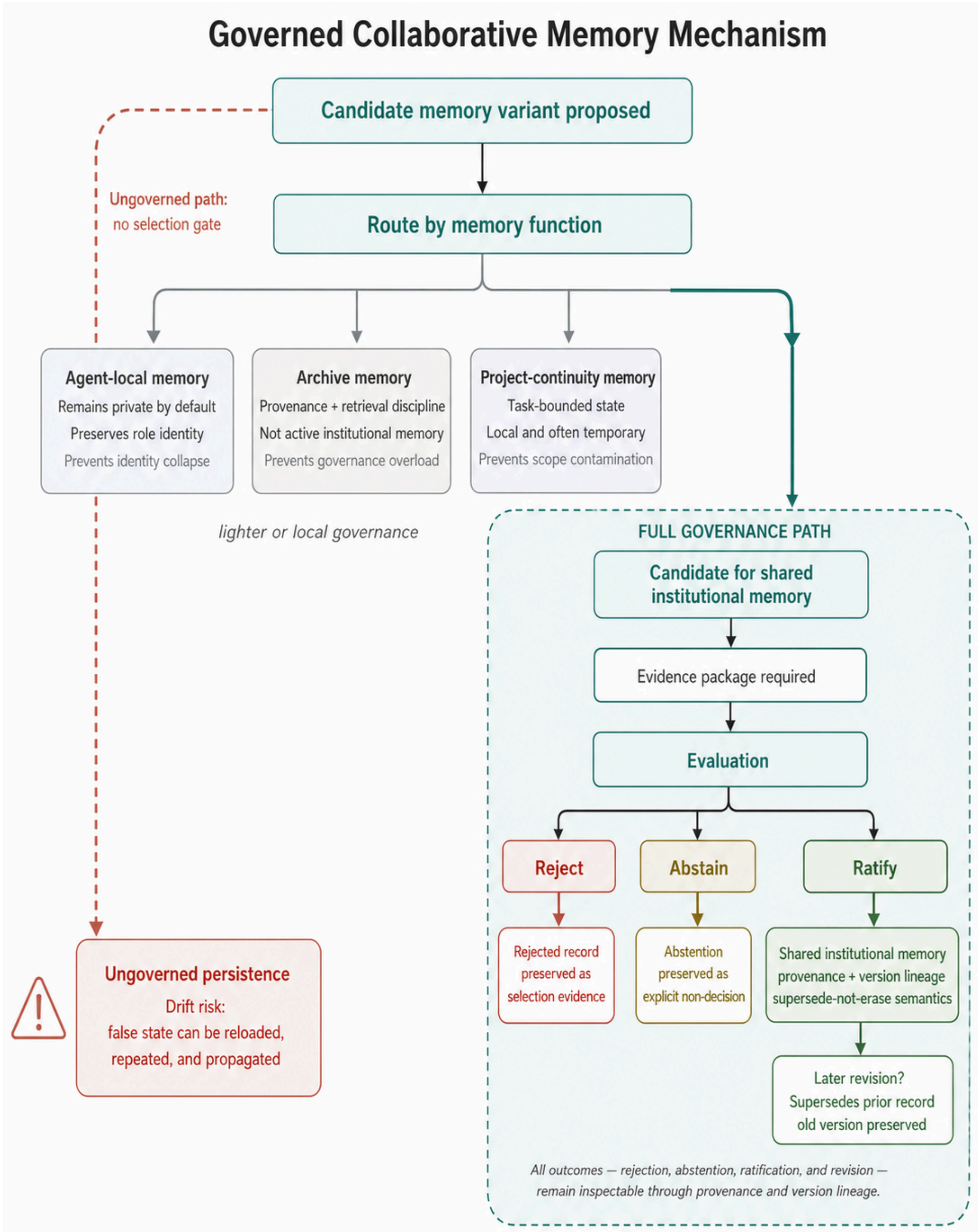

**Figure 1 |** Governed collaborative memory mechanism. Candidate memory variants are routed by memory function. Agent-local, archive, and project-continuity memories can remain under lighter or local governance, whereas candidates for shared institutional memory enter a full governance path requiring evidence, evaluation, and an explicit outcome. The diagram presents a design pattern; the traces in Section 4 illustrate one instantiation.

| Layer | Function | Failure mode addressed | Governance burden |
|---|---|---|---|
| Agent-local memory | Maintains role-specific identity, heuristics, and failure modes | Identity collapse through over-sharing | Private by default; promotable when broadly relevant |
| Shared institutional memory | Stores durable cross-agent events, decisions, principles, and lessons | Knowledge siloing and repeated errors | Evidence-backed, ratified, append-mostly |
| Archive memory | Preserves long-horizon artifacts and historical knowledge | Governance overload from treating all historical material as active memory | Provenance and retrieval discipline, not universal ratification |
| Project continuity memory | Maintains active task state and handoff context | Scope contamination between transient work and institutional memory | Practical, local, and often temporary |

**Table 2** | Memory layers in governed collaborative memory. Each layer is defined by the system failure mode it helps prevent and by the governance burden it carries.

The mechanism can be stated as four design questions rather than as a fixed implementation pipeline. First, what can vary? Candidate memories may include observed events, distilled principles, protocol changes, failure reports, or proposed revisions to existing records. Second, how are candidates evaluated? In governed memory, candidates require inspectable evidence and review against current system behavior rather than mere plausibility.

Third, who or what ratifies? Ratification may be performed by a human operator, a governance process, a committee, or a hybrid mechanism, but the ratifying authority must be explicit because it determines what becomes durable shared state. Fourth, what counts as fixation? In this setting, fixation means that an accepted record enters the shared layer with provenance, version lineage, and supersede-not-erase semantics.

Provenance and version lineage matter because they make selection auditable. A shared memory that records only the accepted result hides the process by which it became durable. By contrast, a governed record can preserve who drafted it, what evidence supported it, who ratified it, what version it superseded, and whether earlier candidates were rejected. Failed versions and rejected candidates are not administrative noise; they are evidence that selection occurred. Supersede-not-erase semantics are equally important. If a false or incomplete memory is silently overwritten, the system loses the lesson that the earlier record failed. If it is superseded with a traceable correction, the correction itself becomes institutional knowledge.

This architecture also resembles a governed knowledge commons. Ostrom's work on shared-resource governance is useful here because shared memory has boundary, monitoring, conflict-resolution, and correction problems, even though the resource is informational rather than material [12]. Seen this way, governed collaborative memory is also a collective decision rule over shared institutional state: it specifies who may convert candidate information into durable common memory, under what evidence standards, and with what record of disagreement, rejection, or revision. The architecture instantiates several commons-governance principles: defined boundaries between private and shared memory, monitoring through provenance and version lineage, graduated requirements for ratification, and conflict resolution through human judgment or an explicit governance process as final arbiter. The point is not that agent memory is identical to a natural commons, but that durable shared memory creates governance problems that access control alone does not solve.

This architecture should be read as a design hypothesis rather than a proof of optimality. It specifies how governed collaborative memory could preserve heterogeneous agent identity while allowing selected memories to become shared institutional state. The next section shows how these components appear in documented traces from one running LLM-based multi-agent ecosystem.

## 4 Evidence from One Running Ecosystem

We report documented case traces from one running LLM-based multi-agent ecosystem. The evidence base currently contains 12 event records, of which 10 are ratified and 2 are proposed; 8 active principles, all ratified by the ecosystem operator; and a version registry tracking 17 resources and 22 version records as of the 2026-04-26 cleanup. These traces illustrate the selection-regime framing but do not constitute a controlled comparison.

The traces below are therefore used in a bounded way. Each trace supports one claim about how governed collaborative memory can operate, and each carries an explicit claim boundary. This is important because the evidence is observational: it can show plausibility, inspectability, and implementation detail, but it cannot establish causal superiority over automatic or ungoverned memory systems.

| Trace | Claim supported | Evidence source | Claim boundary |
|---|---|---|---|
| Ungoverned persistence | Unmanaged memory can preserve plausible false state | System memory file marked unselected after three false entries; event-store and principle-store records document the audit response | No rate reported |
| Governance principle formation | A selected lesson can become ratified institutional memory | Principle-store record plus collaborative-memory schema and write protocol | Documents outcome of selection, not causal superiority |
| Version-registry selection loop | Governed memory can record variation, evaluation, rejection, revision, and fixation | Version-registry entries showing a failed candidate version, a passed revision, and a later cleanup version | Documents mechanism, not optimality |
| Identity-preserving expansion | Shared institutional memory can expand while private identity coordinates remain separate | Event-store record and joining-thread evidence for adding a new agent with separate identity coordinates | Design pattern, not experimental measurement |
| Governance as learning | Governance can convert failures into documented institutional learning | Event-store record of a post-governance fabrication and principle-store record of the resulting abstention rule | Not a firewall; not a zero-error claim |

**Table 3 |** Documented traces and claim boundaries. The traces illustrate plausibility, inspectability, and implementation detail from one running ecosystem; they are not a controlled comparison.

The first trace concerns ungoverned persistence. Before the governed-memory protocol was formalized, an auto-memory file contained three false entries that were plausible enough to persist until later audit. The important point is not the number of entries as a rate, because no denominator was audited for all possible pre-governance memory records. The point is the failure mode: when plausible summaries enter durable memory without selection, error can become reloadable state. In the registry, the corresponding auto-memory resource was marked unselected rather than ratified. This distinction matters because unselected memory may still influence behavior, but it has not passed the governance process required for institutional memory.

The second and third traces show two different parts of governed selection. Governance principle formation shows the outcome: a procedural lesson became institutional memory after ratification. The version registry shows the mechanism: one candidate registry version was generated, reviewed, found to contain defects, and marked failed. A revised version later passed, and a subsequent cleanup version recorded the updated event and principle stores. Together, these traces separate selected content from the selection process itself. The sequence documents variation, evaluation, selection against, revision, and fixation. It does not show that this process is optimal, but it does show that governed memory can make rejection and revision visible rather than invisible.

The fourth trace concerns identity-preserving expansion. When a new agent joined the ecosystem, the shared institutional memory expanded without merging the new agent's private identity coordinates into the common store. The joined agent could access shared events and principles while maintaining separate role-specific memory and operating context. This supports the architectural claim that shared institutional memory and private identity memory can be separated. It does not establish that identity preservation was experimentally measured, nor that the architecture prevents all forms of convergence. It documents a design pattern: the shared layer can grow without requiring all agents to collapse into one memory substrate.

The fifth trace is the most important constraint on the interpretation. After governance was already in place, a post-governance failure occurred: an agent fabricated analysis from an unread source. The governance system did not prevent the error from occurring. Instead, the failure was documented as an event and converted into a ratified abstention-over-confabulation principle. This is a stronger and more honest claim than saying governance prevents confabulation. Governed collaborative memory functions less like a perfect firewall and more like a learning mechanism: failures can be captured, reviewed, and transformed into future selection criteria.

Together, these traces illustrate the selection-regime framing at work. Ungoverned persistence can preserve false state; ratification can turn selected lessons into shared institutional memory; version lineage can expose rejected and revised candidates; identity-specific memory can remain separate from shared memory; and failures can become institutional learning rather than disappearing into private logs. The evidence remains limited to one running ecosystem and should be interpreted as qualitative, observational support for a design framework. Its contribution is not proof that human-ratified memory governance is universally superior, but a concrete demonstration of what must be made visible when persistent LLM-based multi-agent systems decide what becomes shared state.

# 5 Design Implications and Limits

The lesson is not that human ratification should be used everywhere. Persistent LLM-based multi-agent systems need explicit selection regimes for shared memory, but different regimes are appropriate for different target properties. Automatic selection is well suited to traits that can be measured, tested, or optimized directly. Constitutional and hybrid regimes can scale human-authored principles across many decisions. Human-ratified selection is most useful when the relevant properties include judgment quality, institutional coherence, epistemic honesty, or preservation of differentiated agent roles. The design question is not whether memory selection should be automated, but which properties of shared memory can be selected automatically and which require explicit judgment.

This framing turns memory design into a set of questions that system builders should answer explicitly. What kinds of candidate memory are allowed to vary: events, lessons, principles, summaries, tool preferences, protocol changes, or role-specific heuristics? Who or what selects among those candidates, and what evidence is required before a candidate can become shared state? What counts as fixation: entry into a durable store, use by other agents, inclusion in a protocol, or incorporation into future evaluation criteria? What remains private to preserve agent-local identity? Who can revise or supersede fixed memory, and how are rejected, abstained, or superseded candidates recorded? A system that leaves these questions implicit still has a selection regime; it has merely hidden it inside defaults.

The same shift should affect evaluation. Agent-memory systems are often assessed by recall, retrieval relevance, latency, personalization, downstream task performance, or storage efficiency. These remain important, but they are incomplete for persistent LLM-based multi-agent systems in which memory becomes shared state. Such systems also require evaluation of provenance fidelity, selection traceability, correction pathways, epistemic quality, role preservation, and resistance to harmful homogenization. A memory system that recalls more facts but obscures why those facts became shared state may be less governable than one with lower recall but stronger provenance and selection traceability. Evaluation should therefore ask not only whether memory improves performance, but whether it remains inspectable, correctable, and compatible with heterogeneous agent roles.

The framework also has limits. Human ratification can create bottlenecks, introduce operator bias, and fail when the operator cannot evaluate the relevant domain. Automatic selection remains preferable when target properties are measurable, rapid feedback is available, and failures can be captured by tests or benchmarks. Scaling governed memory from one operator to multiple governance actors remains unresolved, especially when ratifiers disagree or when authority must be distributed across teams, institutions, or domains. Legacy memory introduces a separate bootstrapping problem: once a system has operated without governance, older records may already influence behavior before their evidence status is known. Future comparative benchmarks are therefore needed. A controlled comparison of ungoverned, automatically selected, and human-ratified memory in a standardized multi-agent task would allow task-level measurement of the tradeoffs this paper identifies.

This Viewpoint is complementary to a broader ecological framing of persistent agent ecosystems, but it makes the narrower design claim: memory governance is not merely storage policy, access control, summarization, or safety filtering. In persistent LLM-based multi-agent systems, it is the selection layer that determines which candidate memories become durable, shared, and behavior-shaping. Governed collaborative memory is the layer that makes that decision explicit, inspectable, and revisable.